\documentclass[12pt]{l4dc2023}
\newcommand{\R}{\mathcal{R}}
\newcommand{\SA}{\mathcal{S}}
\newcommand{\A}{\mathcal{A}}
\newcommand{\PA}{\mathcal{P}}

\newcommand{\E}{\mathcal{E}}
\usepackage{float}
\usepackage{subfloat}
\usepackage[font=scriptsize]{caption}
\usepackage{latexsym,color}
\usepackage{graphicx}
\usepackage[none]{hyphenat}
\allowdisplaybreaks
\DeclareMathOperator*{\argmax}{arg\,max}
\DeclareMathOperator*{\Argmax}{Arg\,max}

\title[Full Gradient DQN for Average-Reward Criterion]{Full Gradient Deep Reinforcement Learning\\ for Average-Reward Criterion}
\usepackage{times}
\usepackage{lipsum} 
\newcommand\blfootnote[1]{%
  \begingroup
  \renewcommand\thefootnote{}\footnote{#1}%
  \addtocounter{footnote}{-1}%
  \endgroup
}
\usepackage{algpseudocode}
\usepackage{algorithm}
\author{%
 \Name{Tejas Pagare} \Email{tejaspagare2002@gmail.com}\\
 \addr Department of Electrical Engineering, Indian Institute of Technology Bombay, Mumbai 400076, India.
 \AND
 \Name{Vivek Borkar} \Email{borkar.vs@gmail.com}\\
 \addr Department of Electrical Engineering, Indian Institute of Technology Bombay, Mumbai 400076, India.
 \AND
 \Name{Konstantin Avrachenkov} \Email{K.Avrachenkov@inria.fr}\\
 \addr INRIA Sophia Antipolis, 2004, Route des Lucioles, B.P.93, 06902, Valbonne, France.
}

\begin{document}

\maketitle

\begin{abstract}
We extend the provably convergent Full Gradient DQN algorithm for discounted reward Markov decision processes from \cite{FGDQN} to average reward problems. We experimentally compare widely used RVI Q-Learning with recently proposed Differential Q-Learning in the neural function approximation setting with Full Gradient DQN and DQN. We also extend this to learn Whittle indices for Markovian restless multi-armed bandits. We observe a better convergence rate of the proposed Full Gradient variant across different tasks.

 \blfootnote{Additional results, experimental details and analysis can be found in the \href{https://drive.google.com/file/d/1YV9w0-IgEjEZOhmIxIFgewK__tLMazal/view?usp=sharing}{extended appendix}}
\end{abstract}


\section{Introduction}

Average reward Markov Decision Processes (MDPs) are popular stochastic models for the applications 
when the transient behaviour can be ignored and only the long term behaviour matters. 
Average reward MDPs find numerous applications in communication networks \cite{Altman2002},
medical treatment \cite{Schaefer2005} and machine maintenance \cite{Ross2013}.
They are also a good approximation for discounted reward problems with discounts close to one and offer the simplicity of independence from the initial condition in the irreducible MDP case \cite{DPBook}. For these reasons, there is already a substantial body of work on reinforcement learning for this class of problems, e.g., \cite{MahadevanSurvey,DewantoSurvey}. 
In this work, we present off-policy function approximation algorithms for average reward Q-Learning based on RVI Q-learning \sloppy\cite{Abounadi} and Differential Q-Learning \cite{wan2021learning} with DQN \cite{DQN} and Full Gradient DQN which is a variant of DQN proposed and proven to be convergent with the ODE-based analysis in \cite{FGDQN}. 
Well-studied semi-gradient methods for Q-learning with function approximation lack theoretical guarantees and are
shown to diverge \cite{Baird}. The well-known deadly triad of function approximation, bootstrapping, and off-policy training generalizes the divergence issue in RL algorithms \cite{SuttonBook}. For deterministic problems, the full gradient version, which tries to minimize the Bellman error using gradient descent, namely the residual algorithm, is shown to converge in \cite{Baird}. However, for stochastic problems, the proposed residual algorithm considers two independently drawn samples from the simulator for a single update iteration. However, this is not possible in many learning scenario, typically in robotics and video games and is also widely known as `double-sampling' issue.
This is required due to the product of expectations as opposed to the expectation of the product as required by the Bellman error in the update step. Full Gradient DQN, on the other hand, uses the novel experience replay to perform the first expectation approximately ahead of time.
We further extend this to the problem of learning Whittle indices for Restless Multi-Armed Bandits (RMABs) with Q-learning with the average reward criterion, first studied for the tabular case in \cite{WhittleQlearn}. The Whittle index approach allows to deal 
efficiently with many scheduling and resource allocation problems \cite{Whittle,Papadimitriou,Gittins2011MultiArmedBA}. 

The paper is organized as follows. First, we consider a preliminary introduction to average reward Q-learning and propose our Full Gradient DQN (FGDQN) variant. We then propose a variant of Differential Q-learning for FGDQN. Applications to restless bandits are then discussed. Lastly, we show experimental results and conclude with some future directions.

Throughout, for a finite set $S$, we denote by $\PA(S)$ the simplex of probability vectors indexed by the elements of $S$.

\section{Preliminaries}
\label{sec:prelim}
Consider a controlled Markov chain $\{X_n,U_n\}, n \geq 0,$ on a finite state space $\SA, |\SA| = s,$ controlled by a control process $U_n, n \geq 0,$ taking values in a finite action space $\A, |\A| = \ell$, with controlled transition probabilities $p(j|i,u), i, j\in\SA, u \in \A$, satisfying \sloppy $p(j|i,u) \in [0,1] \ \forall \ i,j,u,$ and
$\sum_jp(j|i,u) = 1 \ \forall \ i,u$. Its time evolution is described by
\begin{equation}
\text{Pr}(X_{n+1} = j|X_m,U_m, m \leq n) = p(j|X_n,U_n). \label{controlled}
\end{equation}
Given a per stage reward function $r: \SA\times\A \mapsto \R$, the average reward MDP seeks to maximize the time averaged reward
\begin{equation}
\liminf_{N\uparrow\infty}\mathbb{E}\left[\frac{1}{N}\sum_{n=0}^{N-1}r(X_n,U_n) \right].\label{reward}
\end{equation}
Special classes of interest are stationary policies, each associated with a map $v:\SA \mapsto \A$ such that $U_n = v(X_n) \ \forall n$, and stationary randomized policies, each associated with a map $i \in \SA \mapsto \varphi( \cdot | i) \in \PA(\A)$ such that $P(U_n = u|X_m,U_{m-1}, m \leq n) = \varphi(u|X_n) \ \forall n$. In particular, a stationary randomized policy with $\varphi(v(i)|i) = 1$ for $v: \SA \mapsto \A$ corresponds to a stationary policy $v$. It is clear that $\{X_n\}$ is a Markov chain with transition probabilities $p(j|i,v(i))$, resp.\ $\sum_up(j|i,u)\varphi(u|i)$ under a stationary policy $v$, resp.\ stationary randomized policy $\varphi$. We assume that this chain is irreducible under any $v$, resp.\ $\varphi$. Then it has a unique stationary distribution $\pi_v$, resp.\ $\pi_\varphi$ in $\PA(\SA)$ and (\ref{reward}) a.s.\ equals $\beta_v := \sum_i\pi_v(i)r(i,v(i))$, resp.\ $\beta_\varphi :=\sum_{i,u}\pi_\varphi(i)\varphi(u|i)r(i,u)$. The objective is to maximize (\ref{reward}) over all admissible $\{U_n\}$, i.e., $\{U_n\}$ for which (\ref{controlled}) holds. It is known that an optimal stationary policy exists (\cite{Puterman}, Chapter 8) and is characterized as the maximizer on the right hand side of the dynamic programming equation
\begin{equation}
V(i) = \max_u\Big[r(i,u) - \beta + \sum_jp(j|i,u)V(j)\Big], \ i \in \SA. \label{DP}
\end{equation}
This is an equation in unknowns $V(i), i \in \SA$ and $\beta$.  The $\beta$ is characterized uniquely as the optimal reward $\beta^*$ and $V(\cdot)$ is uniquely characterized modulo an additive constant.

We define the Q-values as the expression in the square brackets on the RHS of (\ref{DP}), i.e.,
$$Q(i,u ) := r(i,u) - \beta + \sum_jp(j|i,u)V(j), \ i\in\SA, u\in\A.$$
Then these satisfy the equation
\begin{equation}
Q(i,u) = r(i,u) - \beta + \sum_jp(j|i,u)\max_vQ(j,v), \ i\in\SA, u\in\A. \label{QDP}
\end{equation}
Again,  $\beta$ is characterized uniquely as the optimal reward $\beta^*$ and $Q(\cdot,\cdot)$ is uniquely characterized modulo an additive constant. If we solve this equation, the optimal stationary policy is given by $v(i) = \argmax Q(i, \cdot )$, breaking ties arbitrarily if non-unique. Furthermore, this does not require the knowledge of the transition probabilities. Also, unlike (\ref{DP}), the nonlinearity, i.e.,\ the `max' operator is now inside the conditional expectation, which facilitates a stochastic approximation version (see, e.g., \cite{BorkarBook}). This is the motivation for using the Q-values as a basis for reinforcement learning algorithms. We do this next.

\section{Full Gradient DQN for average reward}
\label{sec:fgdqn}
A reinforcement learning algorithm to solve (\ref{QDP}) called `RVI Q-learning', based on the classical Relative Value Iteration for solving (\ref{DP}), was proposed and analyzed in \cite{Abounadi} for the tabular case. Here we study a variant based on function approximation using neural networks, that seeks to minimize the so-called `Bellman error'. Thus, with some abuse of notation, let $Q(i,u;\theta)$ denote a parametrized family of approximating functions with parameter $\theta \in \R^d$.  One wants to `learn' a $\theta$ such that
\begin{equation}
Q(i,u;\theta) \approx r(i,u) - \beta + \sum_jp(j|i,u)\max_vQ(j,v;\theta), \ i\in\SA, u\in\A. \label{apprQDP}
\end{equation}
This suggests minimizing the mean square error between the right and left hand sides on , i.e.,
\begin{equation}
\E(\theta) := \mathbb{E}\Big[\Big(r(X_n,U_n) - f(Q;\theta_n) + \sum_jp(j|i,u)\max_vQ(j,v;\theta) - Q(X_n,U_n;\theta)\Big)^2\Big]. \label{Bellman}
\end{equation}
Here $f(Q;\theta_n)$ is the offset that works as a surrogate for $\beta$ as in the RVI Q-learning of \cite{Abounadi}. Since $\beta$ is unknown, we use a surrogate $f(Q;\theta_n)$ by analogy with the original relative value iteration algorithm (see, e.g., \cite{Puterman}). With a suitable choice, this can be shown to converge to $\beta^*$ a.s. Some popular choices are $f(Q) = Q(i_0,u_0)$, $\max_{u}Q(i_0,u),\ \max_{i,u}Q(i,u), \frac{1}{s\ell}\sum_{i,u}Q(i,u)$, for some fixed $i_0 \in \SA, u_0 \in \mathcal{A}$ etc. See \cite{Abounadi}  for a characterization of admissible $f(\cdot)$.

This suggests the naive stochastic (sub)gradient scheme
\begin{eqnarray}
\theta_{n+1} &=& \theta_n + a(n)\left(r(X_n,U_n) - f(Q;\theta_n) + \max_vQ(X_{n+1},v;\theta_n) - Q(X_n,U_n;\theta_n)\right) \nonumber \\
&& \times\left(\nabla_\theta f(Q;\theta_n) - \nabla_\theta Q(X_{n+1},v_{n+1};\theta_n) + \nabla_\theta Q(X_n,U_n;\theta_n)\right). \label{SGD}
\end{eqnarray}
Here $X_{n+1}\sim p(\cdot|X_n,U_n)$, $\{a(n)\}$ is a positive stepsize sequence satisfying the Robbins-Monro conditions $\sum_na(n)=\infty,\ \sum_n a(n)^2<\infty$, $\nabla_\theta$ denotes the gradient with respect to the parameter vector $\theta$ and $v_{n+1}\in\Argmax Q(X_{n+1}, \cdot ; \theta_n)$\footnote{$\Argmax$ is used to denote the possible non-uniqueness of the $\argmax$ operation} which renders the term $\nabla_\theta Q(X_{n+1},v_{n+1};\theta_n)$ a legitimate subgradient by Danskin's theorem \cite{Danskin}.

The catch with this is that if we apply the standard stochastic approximation theory, one sees the right hand side averaged with respect to the conditional distribution, i.e., the transition probability. But then we have a conditional expectation of the product of the two brackets as opposed to a product of their conditional expectations, as suggested by (\ref{Bellman}). One way to avoid this was already proposed in the pioneering work of \cite{Baird}, which is to simulate another random variable $\widetilde{X}_{n+1}$ with exactly the same conditional distribution as $X_{n+1}$ (i.e., $p(\cdot|X_n,U_n)$) and conditionally independent of it given $X_m,U_m, m \leq n$, then replace $X_{n+1}$ in the second bracket by $\widetilde{X}_{n+1}$. But this is an awkward exercise and adds an additional overhead in the simulation, where the Markov chain simulator is often separately available and one does not want to `dig into it'. Hence we use an alternative strategy based on experience replay, by replacing (\ref{SGD}) by
\begin{eqnarray}
\theta_{n+1} &=& \theta_n + a(n)\overline{\Big(r(X_n,U_n) - f(Q;\theta_n) + \max_v Q(X_{n+1},v;\theta_n) - Q(X_n,U_n;\theta_n)\Big)} \nonumber \\
&& \times\Big(\nabla_\theta f(Q;\theta_n) - \nabla_\theta Q(X_{n+1},v_{n+1};\theta_n) + \nabla_\theta Q(X_n,U_n;\theta_n)\Big), \label{ER}
\end{eqnarray}
where the overline denotes empirical average over triplets $(X_m, U_m, X_{m+1}), m < n,$ for which $\{X_m=X_n\ \&\ U_m = U_n\}$. This sets it apart from the classical experience replay (i.e., sampling random transitions), but the advantage is that it achieves the approximate conditional expectation as desired for the first bracket of (\ref{SGD}). That of the second bracket is then taken care of by the averaging effect of stochastic approximation. For deterministic problems, this is equivalent to the expression without the overline, thereby saving the extra computation needed.

We contrast this with what would be the natural extension of the DQN algorithm (\cite{DQN}) for average reward, viz.,

\begin{eqnarray}
\label{eqn:detdqn}
\theta_{n+1} = \theta_n + \dfrac{a(n)}{M}\times \sum_{m=1}^{M}\Big((Z_{n(m)}-Q(X_{n(m)},U_{n(m)};\theta_n))\nabla_\theta Q(X_{n(m)},U_{n(m)};\theta_n)\Big),
\end{eqnarray}
where $(X_{n(m)},U_{n(m)},X_{n(m)+1}), 1\leq m \leq M$ are randomly selected triplets from the experience replay, and
\begin{equation}
    Z_{n(m)} = r(X_{n(m)},U_{n(m)}) +\max_v Q(X_{n(m)+1}, v; \tilde{\theta}_n) -f(Q;\tilde{\theta}_n)
    \label{DQN}
\end{equation}
is the `target' being chased by $Q(X_{n(m)},U_{n(m)})$ and $\tilde{\theta}_n$ is updated on a slower time scale by setting it equal to $\theta_n$ periodically and left unaltered in between. As pointed out in (\cite{FGDQN}), there are several theoretical issues about DQN, but it remains a popular scheme because of good empirical behavior. We shall numerically compare our scheme (\ref{ER}) above with (\ref{DQN}).

We conclude this section with some comments on the convergence analysis, which we do not pursue in detail here because it goes more or less along standard lines. If the averaging due to experience replay is exact, we have an exact stochastic subgradient scheme which under reasonable `richness' condition on the noise, is known to converge a.s.\ to a local minimum (or to a connected set thereof if they are not isolated). The condition on noise requires that it be adequate in all directions, which is easily ensured, e.g., as in \cite{FGDQN}, by adding a small extraneous zero mean noise. But this is rarely required in practice. The errors due to inexact averaging by experience replay etc., if small, will lead to a weaker claim, viz., convergence to a small neighborhood of a local minimum. These claims are standard in stochastic approximation theory, see, e.g., Chapter 11 of \cite{BorkarBook}.

\section{Differential Q-Learning}
\label{sec:diffq}
We modify the Differential Q-learning algorithm  described in \cite{wan2021learning}
for off-policy control in tabular setting to neural function approximation. The idea here is to maintain a scalar proxy $\Bar{R}$ similar to $f(Q)$ in RVI Q-learning which is updated based on the temporal difference error. Following we consider a variant based on Full Gradient DQN. Due to the $\theta$ dependence of $\Bar{R}$, we maintain similar iteration for $\nabla_\theta \Bar{R}$ denoted as $Y_n$ which is used in the full gradient of Q-iteration as follows:
\begin{align}
\theta_{n+1} &= \theta_n-a(n)\overline{\Big(r(X_n,U_n)+\max_{a'}Q(X_{n+1},a';\theta_n)- \bar R_n-Q(X_n,U_n;\theta_n)\Big)}\nonumber\\&\quad\times\Big(\nabla_\theta Q(X_{n+1},v_{n+1};\theta_n)-Y_n-\nabla_\theta Q(X_n,U_n;\theta_n)\Big)\\
      \bar R_{n+1} &=  \bar R_{n} + \dfrac{\eta a(n)}{|\SA||\mathcal{A}|}\times \sum\limits_{s\in \SA,a\in \mathcal{A}} \Big(r(s,a)+\max_{a'}Q(s',a';\theta_n)- \bar R_n-Q(s,a;\theta_n)\Big)\\
      Y_{n+1} &= \nabla_\theta \bar R_{n+1} = Y_n+\dfrac{\eta a(n)}{|\SA||\mathcal{A}|}\times \sum\limits_{s\in \SA,a\in \mathcal{A}} \Big(\nabla_\theta Q(s',v;\theta_n)-Y_n-\nabla_\theta Q(s,a;\theta_n))\Big)
\end{align}
where $s'\sim p(\cdot|s,a),\ v\in\Argmax Q(s',\cdot;\theta_n)$ and $\eta$ is a positive constant which can be thought of as a parameter which controls the speed of the $\Bar{R}$ update w.r.t. $Q$ update. Note that in the DQN variant which considers the semi-gradient in the $Q$-iteration, we won't need $Y_n$ i.e. $Y_n=0\ \forall n$.

\section{Application to restless bandits}
\label{sec:rest}
The problem of Markovian restless bandits is as follows. One has $N>1$ Markov chains $\{X^i_n, n \geq 0\}, 1 \leq i \leq N,$ taking values in discrete state spaces $S^i$ resp. Each has two `modes' of operation, active and passive, with corresponding transition probabilities and per stage rewards given by $p^i_a(j|k), r^i_a(k),  j, k \in S^i,$ for the active mode and  $p^i_b(j|k), r^i_b(k), j, k \in S^i,$ for the passive mode, respectively. The problem is to maximize the average reward
\begin{equation}
\liminf_{n\uparrow\infty}\mathbb{E}\left[\frac{1}{n}\sum_{m=0}^{n-1}\sum_{i=1}^N(\nu_m^ir^i_a(X^i_m) + (1 - \nu^i_m)r^i_b(X^i_m))\right], \label{RB}
\end{equation}
where $\nu^i_m = 1$ if the $i$th chain is in the active mode and zero, otherwise. This is to be done subject to the constraint
\begin{equation}
\sum_{i=1}^N\nu^i_n = M \ \forall  n, \label{constraint}
\end{equation}
for a prescribed $M < N$. In a classic article, Whittle \cite{Whittle} approached this problem, now known to be PSPACE-hard \cite{Papadimitriou}, by first relaxing the per stage constraint (\ref{constraint}) to the average constraint
\begin{equation}
\liminf_{n\uparrow\infty}\mathbb{E}\left[\frac{1}{n}\sum_{m=0}^{n-1}\sum_{i=1}^N\nu^i_m\right] = M, \label{constraint2}
\end{equation}
which makes it a classical constrained MDP with average reward. Using the Lagrange multiplier approach (which can be rigorously justified), it becomes an unconstrained average reward MDP with separable reward and separable constraint function. The Lagrange multiplier then decouples it into separate uncoupled MDPs. Using this as a motivation, Whittle first introduced a subsidy $\lambda$ for passivity added to the reward for being passive, and called the problem (Whittle) indexable if the states in which it is optimal to be passive under subsidy $\lambda$, increases from the empty space to the entire state space as $\lambda$ increases from $-\infty$ to $\infty$. If so, assign to each state $k$, the (Whittle) index $\lambda^*(k)$ defined as that value of the subsidy $\lambda$ for which both active and passive modes are equally desirable. The (Whittle) index policy then at time $n$ is to sort $\lambda^*(X_n)$ in decreasing order (any ties being resolved according to some pre-specified rule) and then render the top $M$ chains active, the rest passive. This is a heuristic that is known to work well in practice and is asymptotically optimal in a certain sense when $N\uparrow\infty$ \cite{Weber,Gittins2011MultiArmedBA}.

The chains are now coupled only through the index policy and the definition of the index for a chain depends on the chain alone. Hence, we drop the superscript $i$ henceforth and look at a single individual chain controlled by a $\{0,1\}$-valued control process $\{\nu_n\}$ so as to maximize the average reward
\begin{equation}
\liminf_{n\uparrow\infty}\mathbb{E}\left[\frac{1}{n}\sum_{m=0}^{n-1}(\nu_m^ir^i_a(X_m) + (1 - \nu_m)r^i_b(X_m))\right]. \label{RB2}
\end{equation}
Let $Q(i,u), i \in \SA, u \in \{0,1\}$, be the Q-values for this constrained problem. It is clear that the condition for the threshold for a switch from passive to active or vice versa at state $\hat{k}$ is given by the equation
$$
Q(\hat{k},1) = Q(\hat{k},0),
$$
where the dependence of $Q(\hat{k},\cdot)$ on $\lambda$ is implicit.  Thus, to obtain the Whittle index for state $\hat{k}$, one needs to solve this simultaneous equation in $Q(\cdot, \cdot)$ and $\lambda$. The solution perforce will depend on the choice of $\hat{k}$. We parameterize $\lambda$ by a parameterized family $(\hat{k},\sigma)\rightarrow \lambda (\hat{k};\sigma)$ and $Q$ by parameterized family $(x,u,\lambda(\hat{k};\sigma),\theta)\rightarrow Q(x,u,\lambda(\hat{k};\sigma);\theta)$ where $x\in\SA$ and $u\in\{0,1\}$. This is done to render the implicit dependence of $Q$ on $\lambda(\hat{k})$ for each $\hat{k}$ separately.  \\
The above equation suggests minimizing the following error
\begin{equation}
\Bar{\E}(\sigma) = \mathbb{E}\Big[ \Big( Q(\hat{k},1,\lambda(\hat{k};\sigma);\theta)-Q(\hat{k},0,\lambda(\hat{k};\sigma);\theta) \Big)^2 \Big].
\end{equation}
The update iteration for $\sigma$ based on the mini-batch stochastic gradient descent is as follows
\begin{eqnarray}
    \sigma_{n+1} = \sigma_n- \dfrac{b(n)}{M}\times\sum_{m=1}^M\nabla_\sigma\Big(Q(X_{n(m)},1,\lambda(X_{n(m)};\sigma_n);\theta_n)-Q(X_{n(m)},0,\lambda(X_{n(m)};\sigma_n);\theta_n)\Big)^2
 \label{eqn:whittle}
\end{eqnarray}
where the stepsizes $\{b(n)\}$ satisfy
$$b(n) > 0, \ \sum_nb(n) = \infty, \ \sum_nb(n)^2 < \infty, \ b(n) = o(a(n)),$$
and the $Q(j,u,\lambda(\hat{k};\sigma);\theta_n), j \in \SA, u \in \{0,1\}$ values are updated \textit{for each $\hat{k}\in \SA$ separately} with modified reward as \\
\[\tilde{r}(X_n,U_n;\hat{k})= (1-U_n)(r_b(X_n)+\lambda(\hat{k};\sigma))+U_nr_a(X_n)\]
using (\ref{ER}) for the FGDQN algorithm and (\ref{DQN}) for the DQN algorithm. Overall, the algorithm can be viewed as two time scale stochastic approximation where the Whittle index is updated on a slower timescale and $Q$-values are updated on a faster time scale. In general one would maintain different $Q$ and $\lambda$ neural networks for different arms, which can be shared for statistically identical arms reducing the search space for $Q$-learning from $(2|\SA|)^N$ to $2|\SA|^2+|\SA|$. 

\section{Experiments}
\label{sec:exps}
\begin{figure}[h]
  \centering
  \begin{minipage}[b]{0.25\textwidth}
    \includegraphics[width=1.25\textwidth]{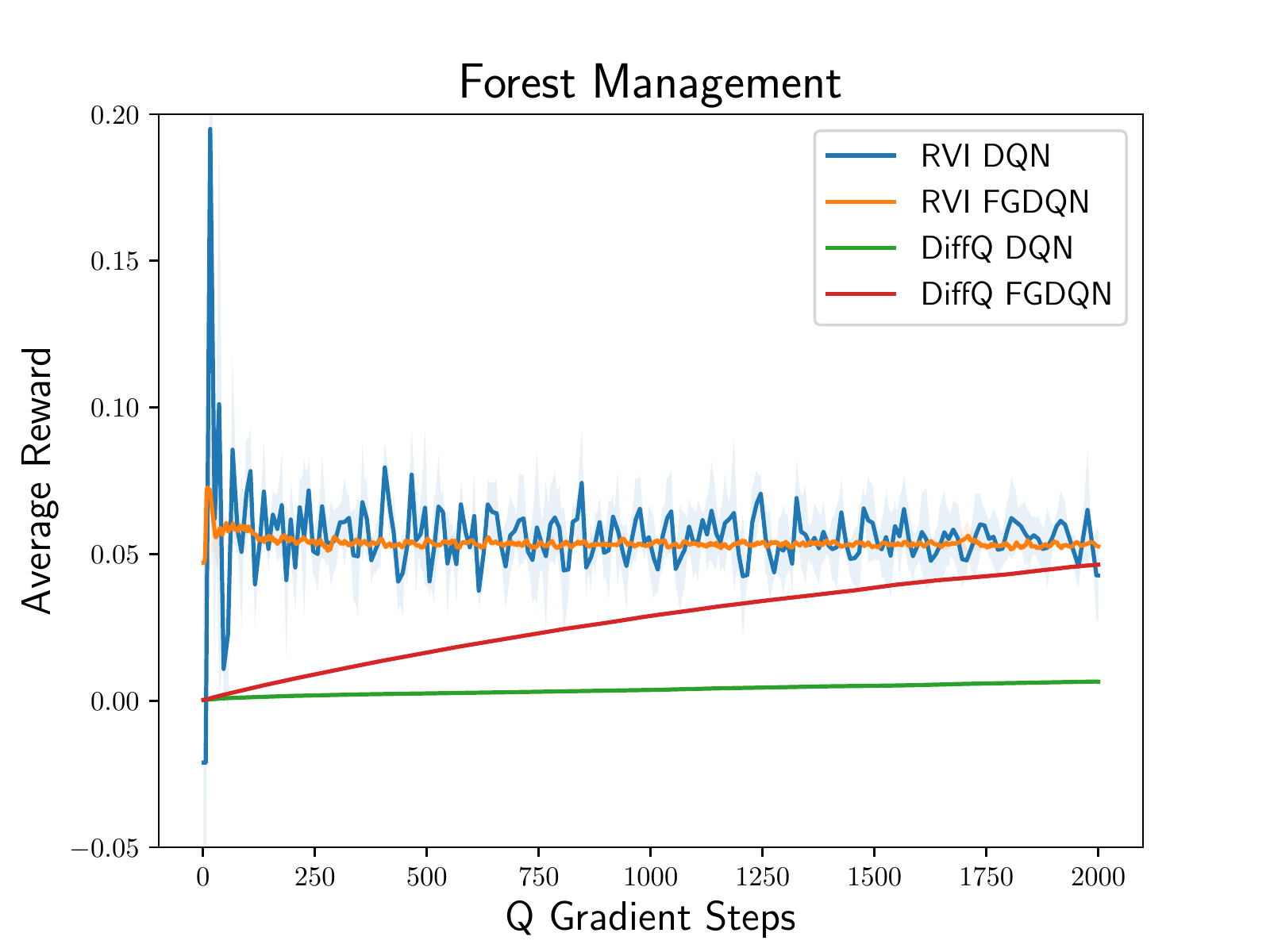}
  \end{minipage}\qquad
  \begin{minipage}[b]{0.25\textwidth}
    \includegraphics[width=1.25\textwidth]{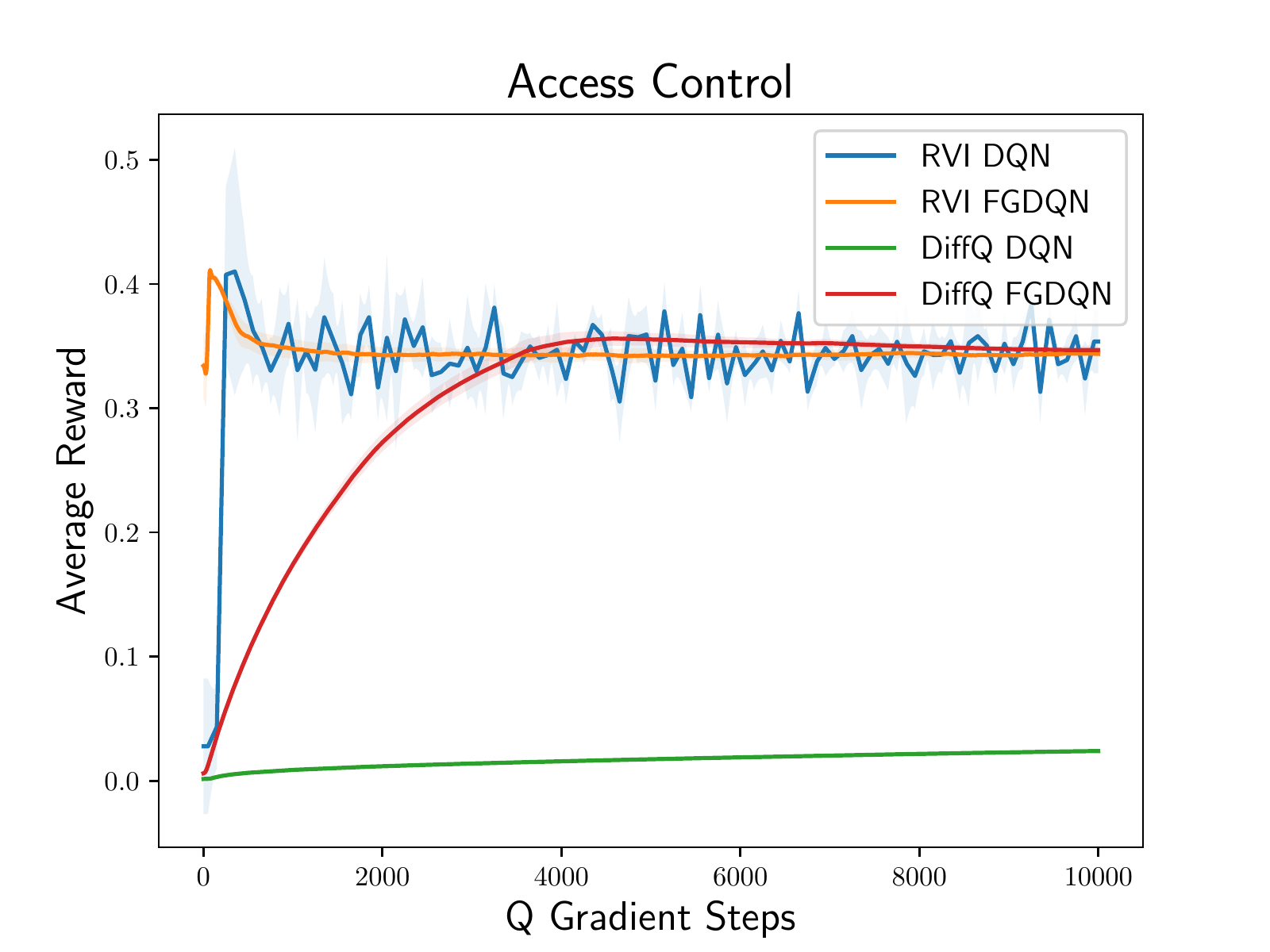}
  \end{minipage}\qquad
  \begin{minipage}[b]{0.25\textwidth}
    \includegraphics[width=1.25\textwidth]{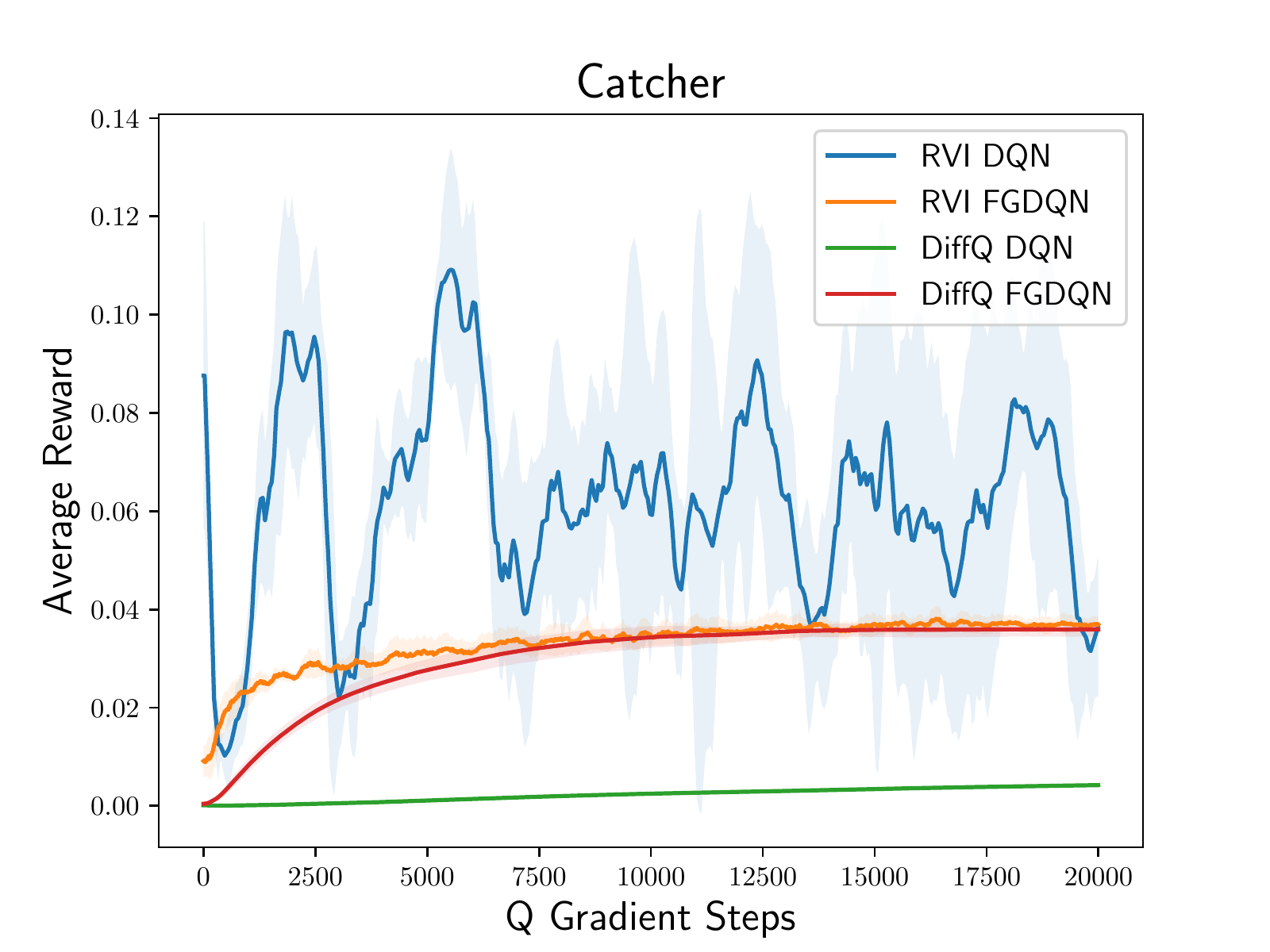}
  \end{minipage}
  \caption{Average Reward Proxy}
  \label{img:proxy}
\end{figure}

In our experiments, we test RVI Q-Learning and Differential Q-Learning with function approximation based on FGDQN and DQN. 
We consider infinite-horizon problems with varying difficulty \footnote{Check \href{https://drive.google.com/file/d/1YV9w0-IgEjEZOhmIxIFgewK__tLMazal/view?usp=sharing}{extended appendix} for problem details.}: Forest Management \cite{forest}, Access Control Queuing \cite{SuttonBook} and Catcher \cite{PLE}. 
We further modify the FGDQN iteration (\ref{ER}) to incorporate the benefit of mini-batch as in DQN to reduce variance as follows:
\begin{eqnarray}
\label{eqn:fgdqn}
\theta_{n+1} &=& \theta_n -\dfrac{a(n)}{M}\times\sum_{m=1}^{M} \Bigg(\overline{\Big(r(X_{n(m)},U_{n(m)}) +\max_v Q(X_{n(m)+1}, v; \theta_n) -f(Q;\theta_n)} \nonumber\\
&& \overline{- Q(X_{n(m)}, U_{n(m)};\theta_n)\Big)}\times \Big(\nabla_\theta Q(X_{n(m)+1}, v_{n(m)+1}; \theta_n) - \nabla_\theta f(Q;\theta_n) -\nonumber \\
&& \nabla_\theta Q(X_{n(m)}, U_{n(m)}; \theta_n)\Big)\Bigg),\ \ n\geq 0,
\end{eqnarray}
where $(X_{n(m)},U_{n(m)},X_{n(m)+1}), 1\leq m \leq M$ are randomly selected triplets from the experience replay, and $v_{n(m)+1}\in \Argmax Q(X_{n(m)+1},\cdot;\theta)$.\\
For FGDQN, we use $f(Q)=Q(s_0,a_0)$ for fixed $s_0\in\SA$ and $a_0\in\mathcal{A}$, obtained as the most frequently visited state-action pair in the replay buffer at the start of the training.
\begin{figure}[h]
  \centering
  \begin{minipage}[b]{0.25\textwidth}
    \includegraphics[width=1.25\textwidth]{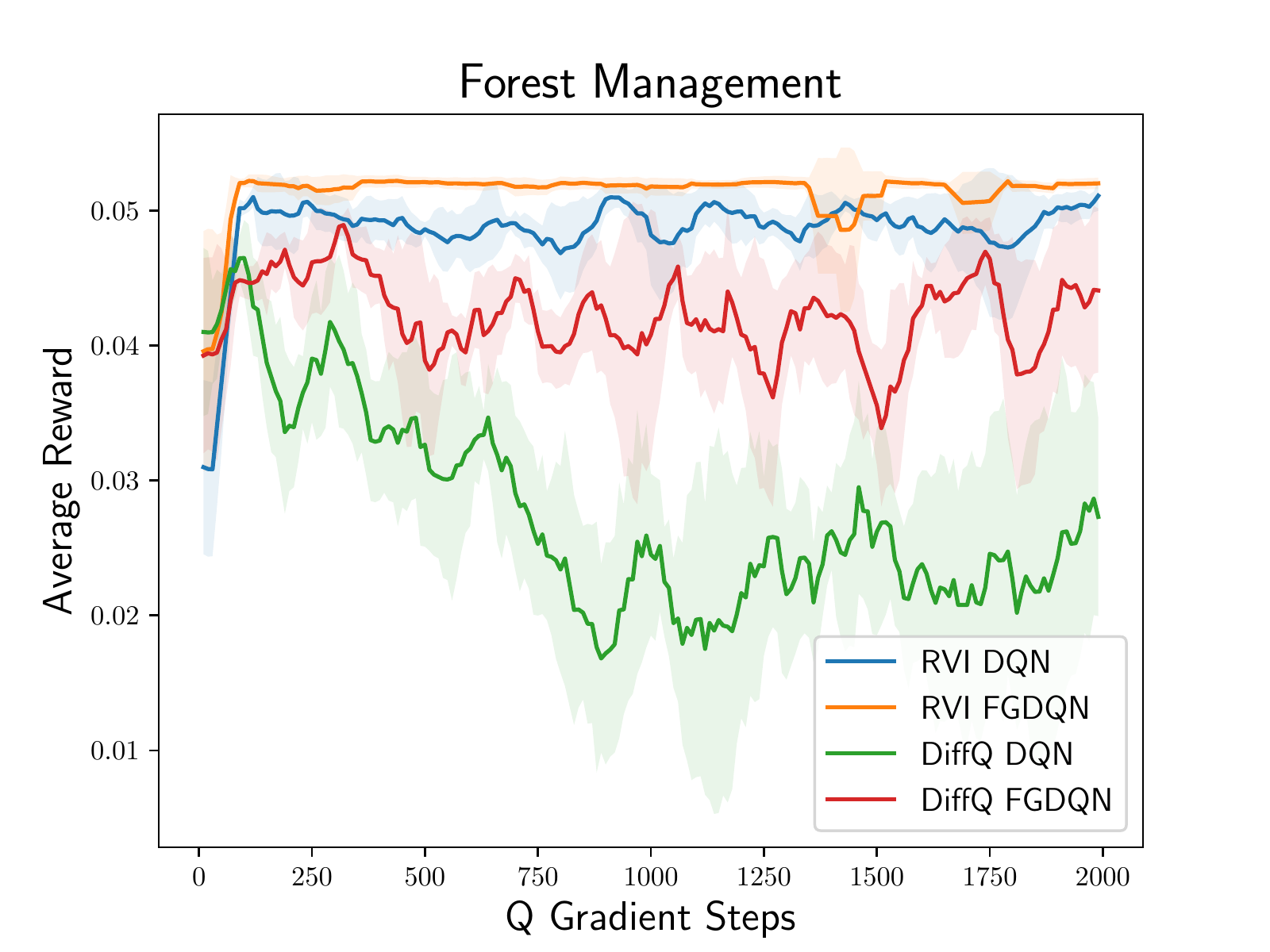}
  \end{minipage}\qquad
  \begin{minipage}[b]{0.25\textwidth}
    \includegraphics[width=1.25\textwidth]{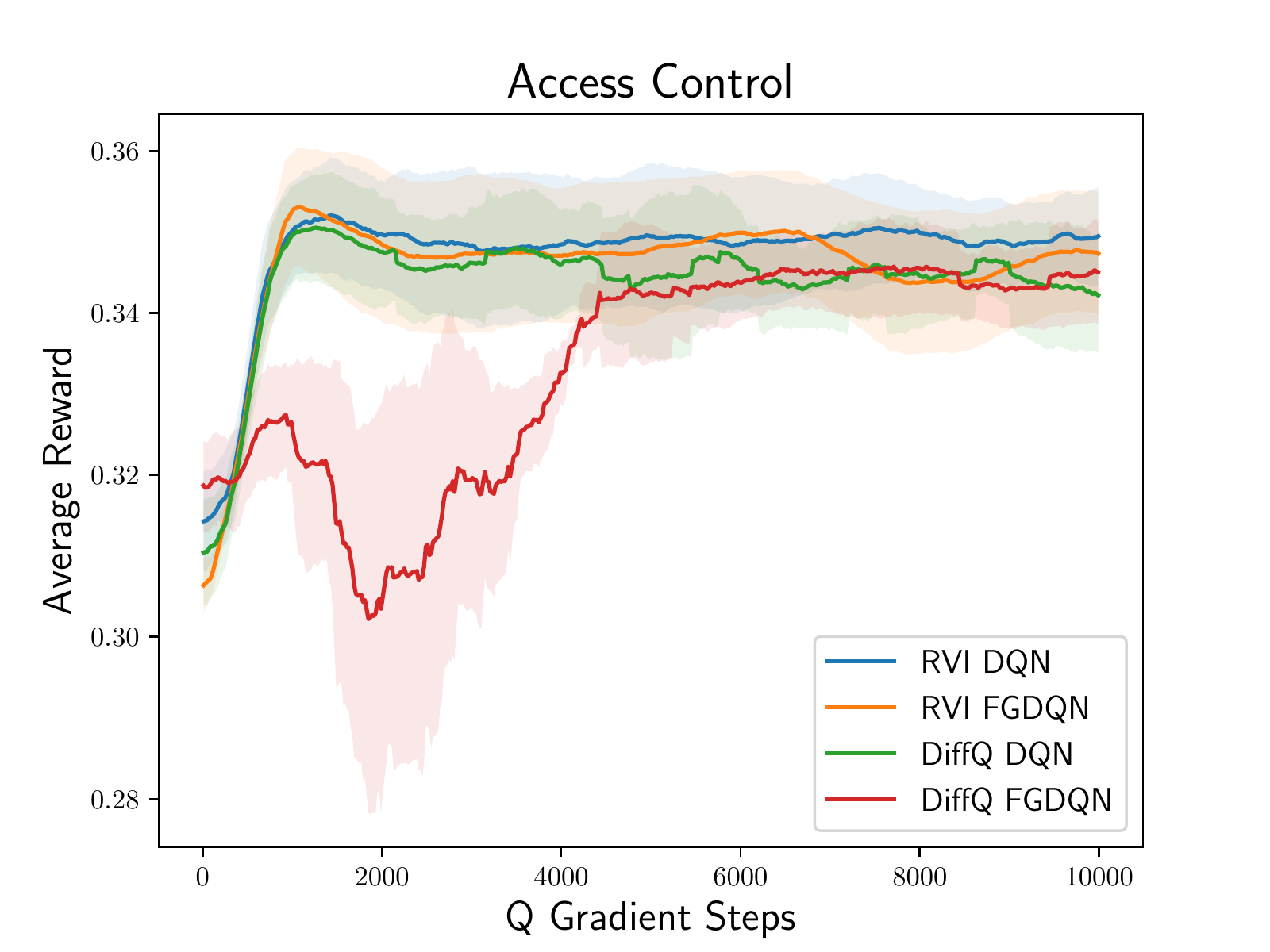}
  \end{minipage}\qquad
  \begin{minipage}[b]{0.25\textwidth}
    \includegraphics[width=1.25\textwidth]{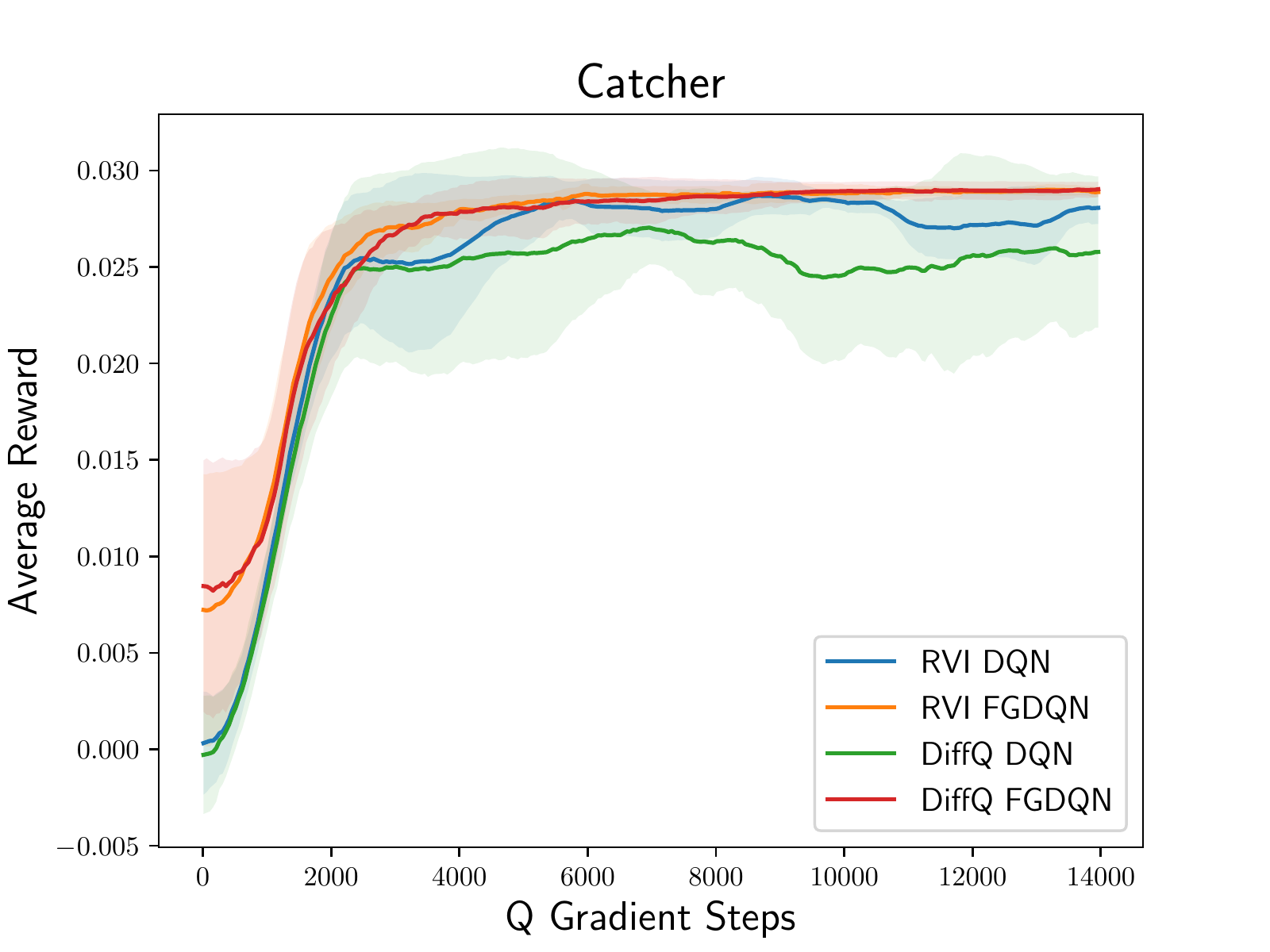}
  \end{minipage}
  \caption{Evaluation Average Reward}
  \label{fig:evalrew}
\end{figure}
Fig. (\ref{fig:evalrew})\footnote{All the figures show data on five different randomly chosen seeds. Further, we show the line plots with moving average (dark)  over a defined window for better analysis and 95\% confidence interval (light).} shows average evaluation reward calculated by running the $Q$-value maximizing policy for 1000 time steps.
In Fig. (\ref{img:proxy})  we plot $Q(s_0,a_0;\tilde{\theta})$\footnote{We observe the zigzag effect in \textsf{{RVI DQN}} since the target network $\tilde{\theta}$ is updated periodically by setting it to $\theta$.}, $Q(s_0,a_0;\theta)$ for \textsf{\small{RVI DQN}} and \textsf{\small{RVI FGDQN}} resp. and $\Bar{R}$ for Differential Q-learning (DiffQ). Asymptotically we want these values to converge to the true average reward $\beta^*$. This convergence can be observed for the \textsf{\small{RVI DQN}},  \textsf{\small{RVI FGDQN}} and \textsf{\small{DiffQ FGDQN}}, but not for \textsf{\small{DiffQ DQN}}, for which $\bar{R}$ remains close to $0$ and hence attributes to relatively poor performance of \textsf{\small{DiffQ DQN}} in the Fig. (\ref{fig:evalrew}). Note in Access Control, \textsf{\small{DiffQ FGDQN}} converges when $\bar{R}$ converges to the true value at about 4000 $Q$-gradient steps. Overall, this shows the direct correlation between the convergence of $\bar{R}$ and the algorithm's performance. In all the experiments, we set $\bar{R}=0$ i.e. $\min r(\cdot,\cdot)$ during initialization which plays a significant role in the convergence rate of the DiffQ algorithm.

\subsection{Restless Bandits}
For Restless Multi-Armed Bandits, we consider the problems of Circulant dynamics \cite{circulant}, Restart problem \cite{WhittleQlearn} and Deadline scheduling \cite{deadline}. We observe unstable learning and thus poor performance of Differential Q-learning hence we don't show them here \footnote{Results can be found in \href{https://drive.google.com/file/d/1YV9w0-IgEjEZOhmIxIFgewK__tLMazal/view?usp=sharing}{extended appendix}}. 
\begin{figure}[!hp]
  \includegraphics[width=1\textwidth]{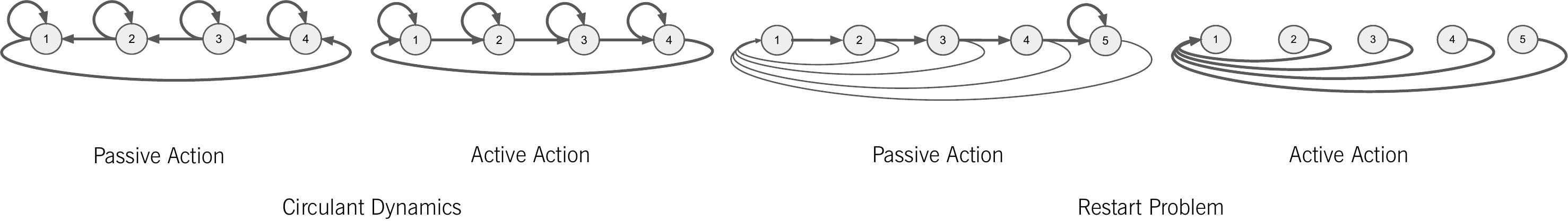}
  \caption{Markov Chains}
  \label{fig:MDP}
\end{figure}

Fig. (\ref{fig:MDP}) shows the Markov chains for active and passive action for both Circulant dynamics and Restart problem. In circulant dynamics problem the state increments (decrements) or remains same with same probability for active (passive) action. Reward function is given as $r(1,a)=-1,\ r(4,a)=1$ and $r(s,a)=0$ for others, independent of the action. Restart problem, restarts the MDP to state $1$ for active action and for passive action, it increments (for $s=5$ it stays at $5$) or restarts to state $1$ with probability 0.9 and 0.1 resp. The reward function $r(s,1)=0.9^s$, and $r(s,0)=0$ for state $s\in \{1,2,3,4,5\}$. 
 
\textbf{Deadline Scheduling}: The problem considers scheduling of jobs depending upon the job arrival, workload, deadline for completion, and the processing cost. One such application can be scheduling of Electrical Vehicles (EV) on a charging station. The charging cost depends on the cost of electricity at the
time of charging, and a penalty is imposed when the service provider is unable to fulfill the request. We consider the problem of EV charging station similar to \cite{NeurWIN}, consisting of $N$ charging stations which can charge $M$ vehicles at any instant of time. For any instant, an EV arrives which declares the amount of charging needed $b$ units and the hard deadline $d$. Action is to charge or not charge the vehicle at every instant which are active and passive actions resp. based upon the deadline $D\in[0,d]$ and load $B\in[0,b]$. Reward is $1-c$ for every unit of charge where $c$ is the cost of electricity. At the deadline if the vehicle is not fully charged, a penalty is received depending upon the amount of charge remaining. After the deadline, new vehicle arrives with probability 0.3. The goal of the station is to maximize the infinite horizon average reward. We characterize a problem with maximum load $B_{\max}$ and deadline $D_{\max}$ possible, which leads to problem with state space of size
$(B_{\max}+1)\times(D_{\max}+1)$. We consider two variants, first with $D_{\max}=12, B_{\max}=9$ and $D_{\max}=50, B_{\max}=45$ with 130 and $2346$ number of states resp.

\begin{figure}[!hp]
  \includegraphics[width=1\textwidth]{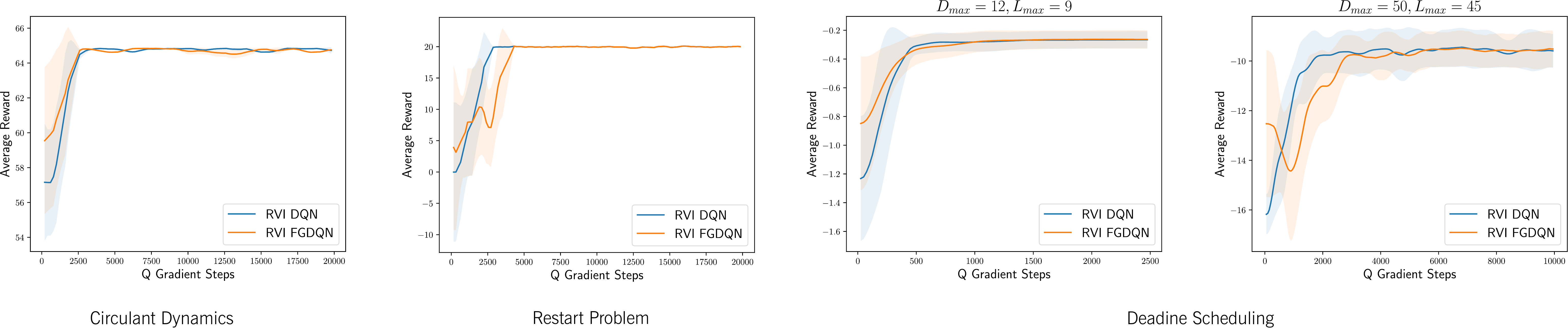}
  \caption{Evaluation Average Reward}
  \label{img:DS}
\end{figure}
For all the problems we consider the restless bandit scenerio with $N=100, M=20$ for circular and restart problems and $N=4, M=1$ for deadline scheduling problem. 
Here we consider the shared architecture, which implies using the same neural network for $Q$-values and whittle index $\lambda$ for all arms. This is done since all the arms are statistically identical (i.e. have same transition matrix and reward function). Fig. (\ref{img:DS}) shows average evaluation reward calculated by running the index policy for 5000 time steps for deadline scheduling with $D_{\max}=50, B_{\max}=45$  and 1000 time steps for others. We consider $f(Q)=Q(s_0,u_0)$ for circular and restart problems and  $f(Q)=\max_u Q(s_0,u)$ for deadline scheduling where $s_0,u_0$ are fixed state and action, which we found to be stable in the respective problems. It can be observed that overall FGDQN starts off quickly compared to DQN and has better convergence rate than DQN for 2 out of 4 problems. 

\section{Conclusion}
We observe that the introduced full gradient variant of DQN (FGDQN) for average reward criterion, at the expense of extra computation, was able to outperform DQN on some tasks for both RVI Q-learning and Differential Q-learning algorithm. Further, we presented an elegant way of solving the problem of restless bandits based on Q-learning in case of complex problems. There are promising research directions where one could consider risk sensitive control with Q-learning \cite{risk} which has applications in finance for portfolio optimization \cite{Bielecki1999}. 
One can also try to create a standard benchmark for average reward criterion similar to Atari \cite{atari} for discounted problems to boost the research in this direction. 
\section{Appendix - Pseudocode}
\begin{equation}
    \label{eqn:finalfgdqn}
\theta_{n+1} = \theta_n -\dfrac{a(n)}{\mathcal{B}}\times\sum_{b=1}^{\mathcal{B}} \Bigg({E_b} \times \Big(\nabla_\theta Z_b- \nabla_\theta Q(x_{n(b)}, u_{n(b)}; \theta_n)\Big)\Bigg),\ \ n\geq 0,
\end{equation}
Problems where state space is huge, iterating over all $\hat{k}\in\SA$ is time consuming hence we instead sample a batch of reference states to update $Q$-values and Whittle index $\lambda$.
\begin{algorithm}[H]\label{algo2}
\SetKwInOut{Input}{Input}
\SetKwInOut{Output}{Output}
\SetAlgoLined
 \textbf{Input:} replay memory $\mathcal{D}$ of size $M$, batch size $\mathcal{B}$, exploration probability $\epsilon$, total gradient steps $T$.\\
 Initialise the weights $\theta \ \& \ \sigma$ randomly for the Q-Network and Whittle-Network. Consider RMAB problem with $N$ statistically identical chains such that at every time step $N$ of them are active. Denote state of the system at time $n$ as $X(n) = (x_1(n),\ldots,x_N(n))$ where $x_i(n)\in\{1,\ldots,d\}$ is a state of chain for $i\in\{1,\ldots,N\}$. Similarly, denote reward and action vectors as $R(n)$ and $U(n)$. \\
\For{n = $1$ to $T$}{
\eIf{\text{Uni}[0,1] $< \epsilon$}{Select $U(n) = (u_1(n),\ldots,u_N(n))$ at random such that $\sum_{i=1}^N u_i(n) = M$}
{Select $u_i(n) = 1$ for largest $N$ whittle indices $\lambda(u_i(n);\sigma)$ }
Add transitions $\{x_i(n),u_i(n),r_i(n),x'_i(n) \} \ \forall i\in\{1,\ldots,N\}$  in $\mathcal{D}$.\\
  Sample random batch of size $\mathcal{B}$ transitions from $\mathcal{D}$.\\
   Sample $\mathcal{B}$ number of reference states from $\mathcal{D}$.\\
    \For{ $b= 1$ to $\mathcal{B}$}{
        \textcolor{blue}{Sample transitions with fixed state-action pair as $x_b,u_b$.\\
        $\text{Set}\ E_b = \overline{ r_b + (1-u_b)\lambda(\hat{k}_b;\theta)+\max_v Q(x'_b,v,\lambda(\hat{k}_b);\theta) - f(Q(\lambda(\hat{k}_b;\theta)))}\\
        \phantom{\text{Set}\ E_b = }\  \overline{-Q(x_b,u_b,\lambda(\hat{k}_b);\theta)}$}\\
        $\text{Set}\ Z_b =  r_b + (1-u_b)\lambda(\hat{k}_b;\theta)+\max_v Q(x'_b,v,\lambda(\hat{k}_b);\theta) - f(Q(\lambda(\hat{k}_b;\theta)))$
        }
    Compute gradient for the average loss over the batch $\mathcal{B}$ and using Eq. (\ref{eqn:finalfgdqn}) update parameters $\theta$.\\\\
  On slower time-scale \textbf{do}\\
 Sample $\mathcal{B}$ number of reference states from $\mathcal{D}$.\\
Perform mini-batch stochastic gradient descent for following mean-squared loss to update $\sigma$ using Eq. (\ref{eqn:whittle})
    \[\mathcal{\Bar{E}}(\sigma)=\lVert Q(\hat{k},1,\lambda(\hat{k};\sigma);\theta)-Q(\hat{k},0,\lambda(\hat{k};\sigma);\theta)\rVert^2\]
  
 }
\caption{Whittle Indices with FGDQN (Statistically Identical Arms)}
\label{fgdqn_whittle_algorithm}
\end{algorithm}
\noindent\textcolor{blue}{Step} is required to approximate the first expectation in the bellman gradient step. For continuous state space, this step can be skipped since only on rare cases we may encounter same states and hence same state-action pair. \\
\acks{Computing resources
were provided by Compute Canada. VB's work was supported in part by a S.\ S.\ Bhatnagar Fellowship from the Government of India.}
\bibliography{Ref}
\end{document}